\documentclass[12pt,preprint,useAMS,usenatbib]{emulateapj}
\usepackage{graphicx}
\usepackage{amssymb}
\usepackage{amsmath}
\usepackage{threeparttable}

\def\msun{\,{M_\odot}}

\def\spose#1{\hbox to 0pt{#1\hss}}
\def\lta{\mathrel{\spose{\lower 3pt\hbox{$\mathchar"218$}}
     \raise 2.0pt\hbox{$\mathchar"13C$}}}
\def\gta{\mathrel{\spose{\lower 3pt\hbox{$\mathchar"218$}}
     \raise 2.0pt\hbox{$\mathchar"13E$}}}

\newcommand{\etal}{{et al.\ }}

\def\kms{\,{\rm km\,s^{-1}}}

\lefthead{Rashkov \etal}
\righthead{}
\submitted{}


\begin{document}

\title{A ``LIGHT," CENTRALLY-CONCENTRATED MILKY WAY HALO?}
\author{Valery Rashkov$^1$, Annalisa Pillepich$^1$, Alis J. Deason$^{1,5}$, Piero Madau$^1$, Constance M. Rockosi$^2$, Javiera Guedes$^3$, and 
Lucio Mayer$^4$}
\altaffiltext{1}{Department of Astronomy and Astrophysics, University of California, Santa Cruz, CA 95064, USA}
\altaffiltext{2}{Department of Astronomy and Astrophysics, UCO/Lock Observatory, University of California, Santa Cruz, CA 95064, USA}
\altaffiltext{3}{Institute for Astronomy, ETH Zurich, 8093 Zurich, Switzerland}
\altaffiltext{4}{Institute of Theoretical Physics, University of Zurich, CH-9057 Zurich, Switzerland}
\altaffiltext{5}{Hubble Fellow.}
\begin{abstract} 
We discuss a novel approach to ``weighing'' the Milky Way (MW) dark matter halo, one that combines the latest samples of halo stars selected from the Sloan Digital Sky Survey (SDSS) with 
state-of-the-art numerical simulations of MW analogs. The fully cosmological runs employed in the present study include ``Eris'', one of the highest-resolution 
hydrodynamical simulations of the formation of a $M_{\rm vir}=8\times 10^{11}\msun$ late-type spiral, and the dark-matter only $M_{\rm vir}=1.7\times 10^{12}\msun$ ``Via 
Lactea II" (VLII) simulation. Eris provides an excellent laboratory for creating mock SDSS samples of tracer halo stars, and we successfully compare their density, velocity anisotropy, 
and radial velocity dispersion profiles with the observational data. Most mock SDSS realizations show the same ``cold veil'' recently observed in the distant stellar halo of the 
MW, with tracers as cold as $\sigma_{\rm los}\approx 50\,\kms$ between 100-150 kpc. Controlled experiments based on the integration of the spherical Jeans equation 
as well as a particle tagging technique applied to VLII show that a ``heavy" $M_{\rm vir}\approx 2\times 10^{12}\msun$ realistic host produces a poor fit to the
kinematic SDSS data. We argue that these results offer added evidence for a ``light,'' centrally-concentrated MW halo.
\end{abstract}

\section{Introduction} \label{intro}

The total mass of the Milky Way (MW) is a key astrophysical quantity for placing our Galaxy in a cosmological context, modeling the formation of the Local Group, deciphering the structure and dynamics of the luminous Galactic components, and ultimately for understanding the complex processes of baryon dissipation and substructure survival within the cold dark matter (CDM) hierarchy. It is, unfortunately, amongst the most poorly known of all Galactic parameters. A common technique to measure the total mass of the MW is to use kinematic tracers such as globular clusters, distant halo stars and satellite galaxies \citep{wil99,xue08,gne10,dea12b,boy13}. In spherical geometry, the Jeans equation provides a link between the host potential of a pressure-supported system and the distribution function of the tracer population in equilibrium with it according to
\begin{equation}
M(<r)={r\sigma_r^2\over G}\left(-{d\ln \rho_*\over d\ln r}-{d\ln \sigma_r^2\over d\ln r}-2\beta\right),
\end{equation}
where $M(<r)$ is the total gravitating mass, $\rho_*$ is the density profile of the tracer stars, $\sigma_r$ is their radial velocity dispersion, and $\beta=1-\sigma^2_t/2\sigma^2_r$ is the velocity anisotropy between the tangential, $\sigma^2_t\equiv \sigma^2_\theta+\sigma^2_\phi$, and radial components of the dispersion. Even with high-quality data (accurate distances and radial velocities), the orbital eccentricities and the density profile of the tracer population are poorly constrained. Without firm knowledge of the tracer properties, Galactic mass measurements suffer from the well-known mass-anisotropy-density degeneracy. The unknown properties and low number of tracers, the uncertainties as to whether some satellites are bound or unbound, and the problematic extrapolations from the inner halo to the virial radius, are all reflected in the wide range of estimates for the virial mass of the MW, $5\times 10^{11}\,\msun<M_{\rm vir}<3\times 10^{12}\,\msun$ \citep[e.g.,][]{wat10}. Note that commonly used mass estimators implicitly assume that kinematic tracers are relaxed. In fact, the much longer dynamical timescales at large radii suggest that the outer reaches of the stellar halo are dominated by unrelaxed substructure. 

In this Letter, we take a different approach to ``weighing'' the MW halo, one that combines the latest samples of halo stars selected from the {Sloan Digital Sky Survey} (SDSS) with 
measured radial velocities to state-of-the-art numerical simulations of realistic MW analogs. The fully cosmological simulations employed in the present study include Eris, one of the 
highest-resolution hydrodynamical simulations ever run of the formation of a $M_{\rm vir}=8\times 10^{11}\msun$ MW-sized disk galaxy \citep{gue11}, as well as the more massive, 
$M_{\rm vir}=1.7\times 10^{12}\msun$, dark-matter only  Via Lactea II (VLII) simulation \citep{die08}. Eris is the first simulated galaxy in a $\Lambda$CDM cosmology in which the galaxy structural properties, mass budget in the various components, and scaling relations between mass and luminosity all appear to be consistent with a host of MW observational constraints. It thus provides an excellent laboratory for creating mock SDSS samples of tracer halo stars 
and compare their density, velocity anisotropy, and radial velocity dispersion profiles with the data. We will show below that Eris' stellar halo reproduces the kinematic 
properties of SDSS halo stars remarkably well, and argue that this offers added evidence for a ``light,'' centrally concentrated MW dark matter halo.     


\section{Halo Tracer Population}
\label{mass_tracers}

To derive precision constraints on the mass of the MW's dark matter halo, we use the samples of blue horizontal branch (BHB) stars specifically targeted by the SDSS and SEGUE for spectroscopy. BHB stars are excellent tracers of Galactic halo dynamics because they are luminous and have a nearly constant absolute magnitude within a restricted color range. The \citet{xue11} sample of BHB stars includes 2,558 stars with galactocentric distances $r<60$ kpc. Their selection combines a photometric color cut based on the analysis by \citet{yan00} and two spectroscopic Balmer line cuts meant to remove contaminating blue straggler (BS) and warm main-sequence stars. With the aim of finding more distant, $r>80$ kpc stars, \citet{dea12b} first implemented a photometric color cut to select objects with $20<g<22$ consistent with faint stars. Follow-up spectroscopy with the VLT-FORS2 revealed 7 BHB and 31 BS stars reaching galactocentric distances of 150 kpc. These were complemented with a sample of N-type carbon (CN) stars \citep{tot98,tot00,mau05,mau08}, and two additional BHB stars \citep{cle05}. The final distant halo star sample comprises 144 BHB and BS stars between 50 and 90 kpc and 17 BHB and CN stars between 80 and 160 kpc. We adopt the binning of all these data exactly as published in Figure 9 of \citet{dea12b}: the radial bins contain as many as 1239 (inner black point) and as few as 9 (outer blue point) stars.   

\section{The Eris Simulation}
\label{mass_simulations}

Details of the Eris simulation are given in \citet{gue11,gue13} and are quickly summarized here for completeness. Eris follows the formation of a $M_{\rm vir}=8\times 10^{11}\msun$ galaxy halo from $z=90$ to the present epoch using the $N$-body + smoothed particles hydrodynamics code {\tt GASOLINE} \citep{wad04}. 
The target halo was selected to have a quiet merger history, and the high resolution region was resampled with 13 million dark matter particles and an equal number of gas particles, for a mass 
resolution of $m_{\rm DM}=9.8\times 10^4\,\msun$ and $m_{\rm SPH}=2\times 10^4\,\msun$, and a gravitational softening of 120 pc (physical). 
Star particles form in cold gas that reaches a density threshold of five atoms cm$^{-3}$, 
supernova explosions deposit an energy of $8\times 10^{50}\,$erg and metals into a ``blastwave radius," and the heated gas has its cooling shut off following \citet{sti06}. 
At $z=0$, Eris is a late-type spiral galaxy of virial radius $R_{\rm vir}=239$ kpc (defined as the radius enclosing a mean density of $93\rho_{\rm crit}$) and 
total stellar mass $M_*=3.9\times 10^{10}\,\msun$. Its rotation curve has a value at 8 kpc (the solar 
circle) of $V_{c,\odot}=206\,\kms$, in good agreement with the recent determination of the local circular velocity, $V_{c,\odot}=218\pm 6\,\kms$, by \citet{bov12}. The best Navarro-Frenk-White (NFW) fit to the 
dark matter profile over the outer region $r/R_{\rm vir}>0.01$ kpc is characterized by a large halo concentration parameter $c\equiv R_{\rm vir}/r_s=24$ as Eris' halo formed early and contracted 
in response to the condensation of baryons in its center.
Mock $i-$band images show that Eris has an $i$-band absolute magnitude of $M_i=-21.7$, an extended stellar disk of exponential scale length $R_d=2.5$ kpc, 
and a bulge-to-disk ratio B/D=0.35. Eris falls on the Tully-Fisher relation and on the stellar mass-halo mass relation at $z=0$. The predicted correlations of stellar 
age with spatial and kinematic structure are in good qualitative agreement with the 
correlations observed for mono-abundance stellar populations in the MW \citep{bir13}. 

\begin{figure}
\centering
\includegraphics[width=0.47\textwidth]{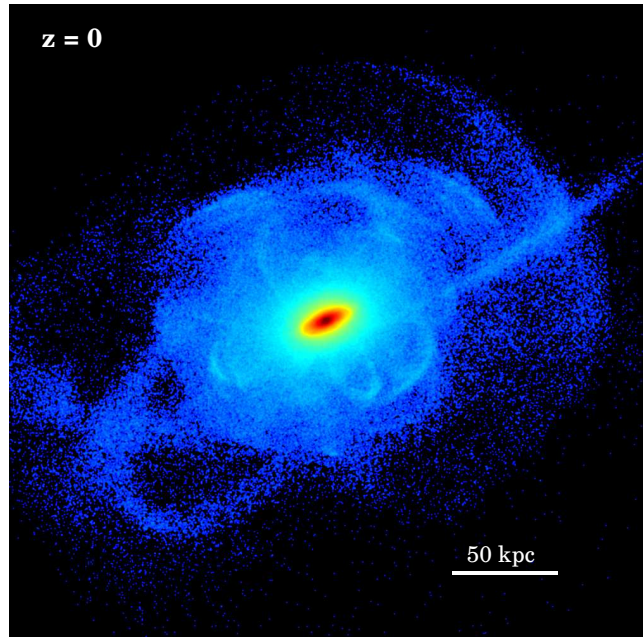}
\vspace{+0.cm}
\caption{\footnotesize External view of Eris' present-day stellar halo (excluding bound satellites). A 300 kpc region centered on the galactic stellar disk is shown in projection. 
The surface mass density ranges from 300 to $10^8$ $M_{\odot}$ kpc$^{-2}$. 
}
\label{fig1}
\vspace{+0.cm}
\end{figure}

Figure \ref{fig1} shows the $z=0$ stellar halo of Eris (excluding satellites) as viewed by an external observer. A variety of substructure, streams and shells, remnants of past accretion 
events, is visible in the image. The mass of the stellar halo (defined by all stars beyond $r=20$ kpc and within $R_{\rm vir}$, and excluding satellites) is $M_*=1.3\times 10^9\,\msun$, 
in good agreement with observational estimates of the MW stellar halo \citep[$M_* \sim 10^{9}$, ][]{mor93, bel08, dea11b}. More than 90\% of these stars formed ``ex-situ", i.e. in 
some satellite progenitor system, and were subsequently accreted in stellar form by the parent galaxy (A. Pillepich et al. 2013, in preparation). 
The outer halo is old and unrelaxed, with a mean stellar age of 10 Gyr. 

\section{Profile and Kinematics of Eris' Stellar Halo}
\label{Eris_kinematics}

\begin{figure}
\centering
\includegraphics[width=0.47\textwidth]{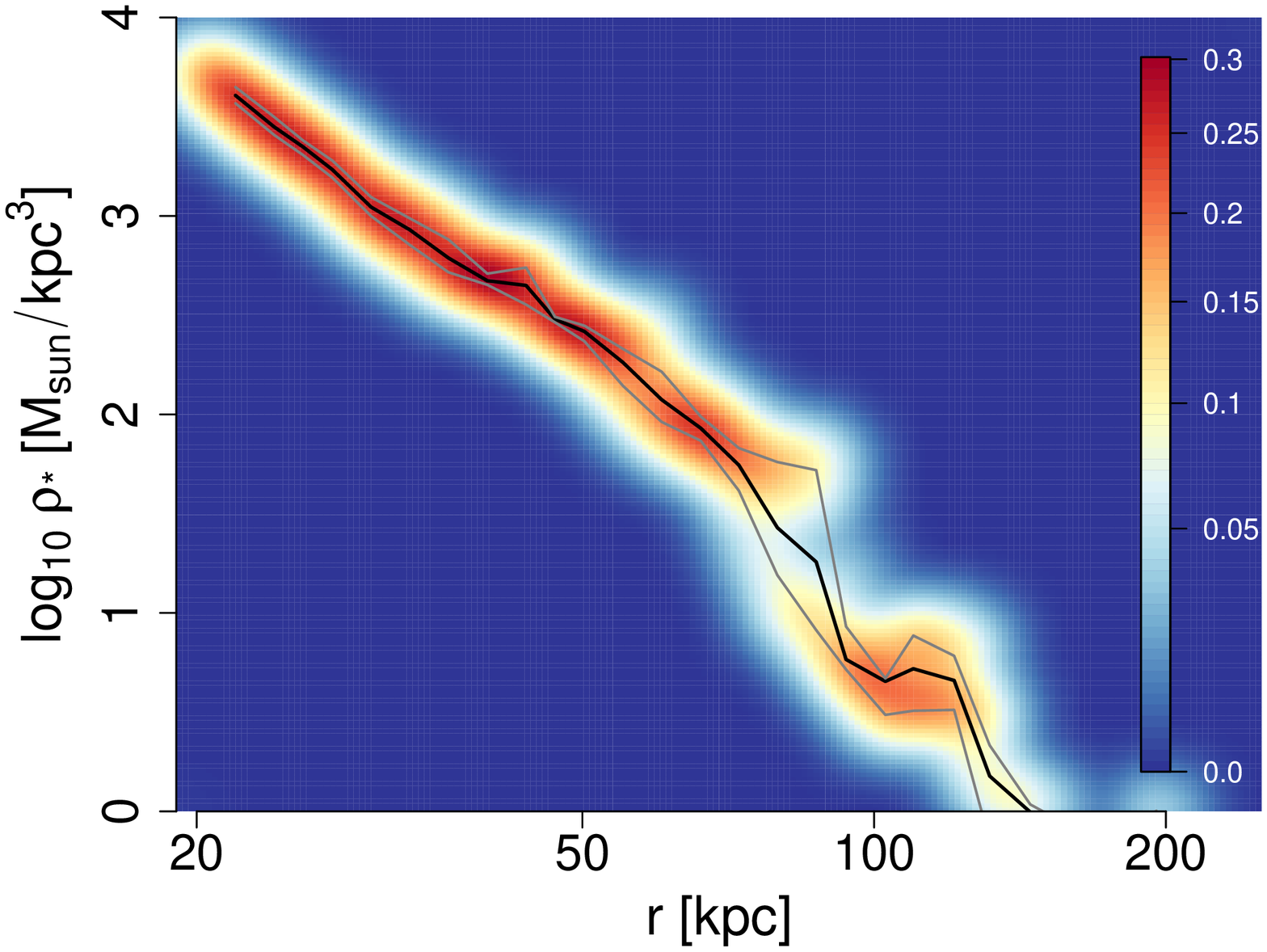}
\includegraphics[width=0.47\textwidth]{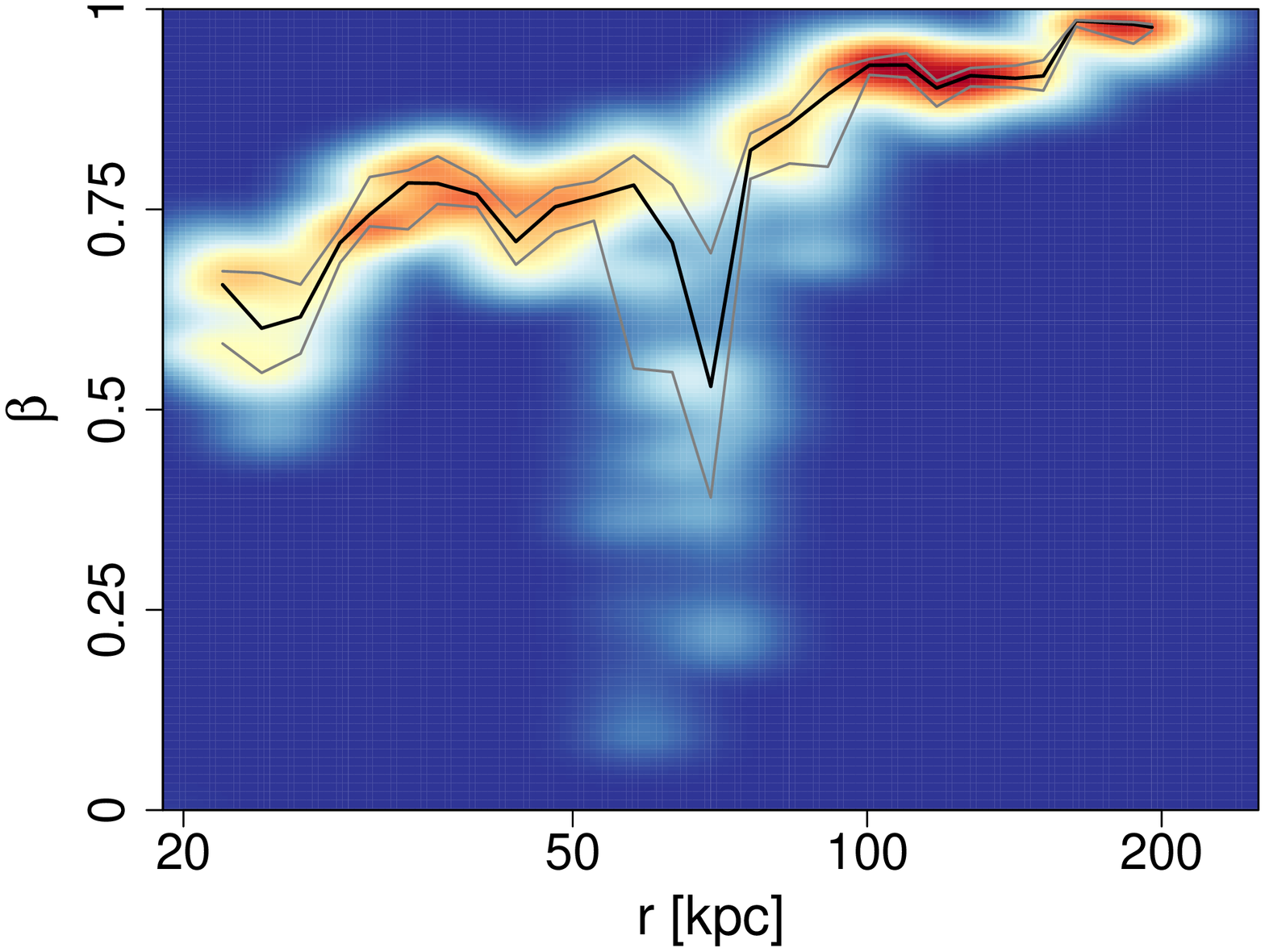}
\includegraphics[width=0.47\textwidth]{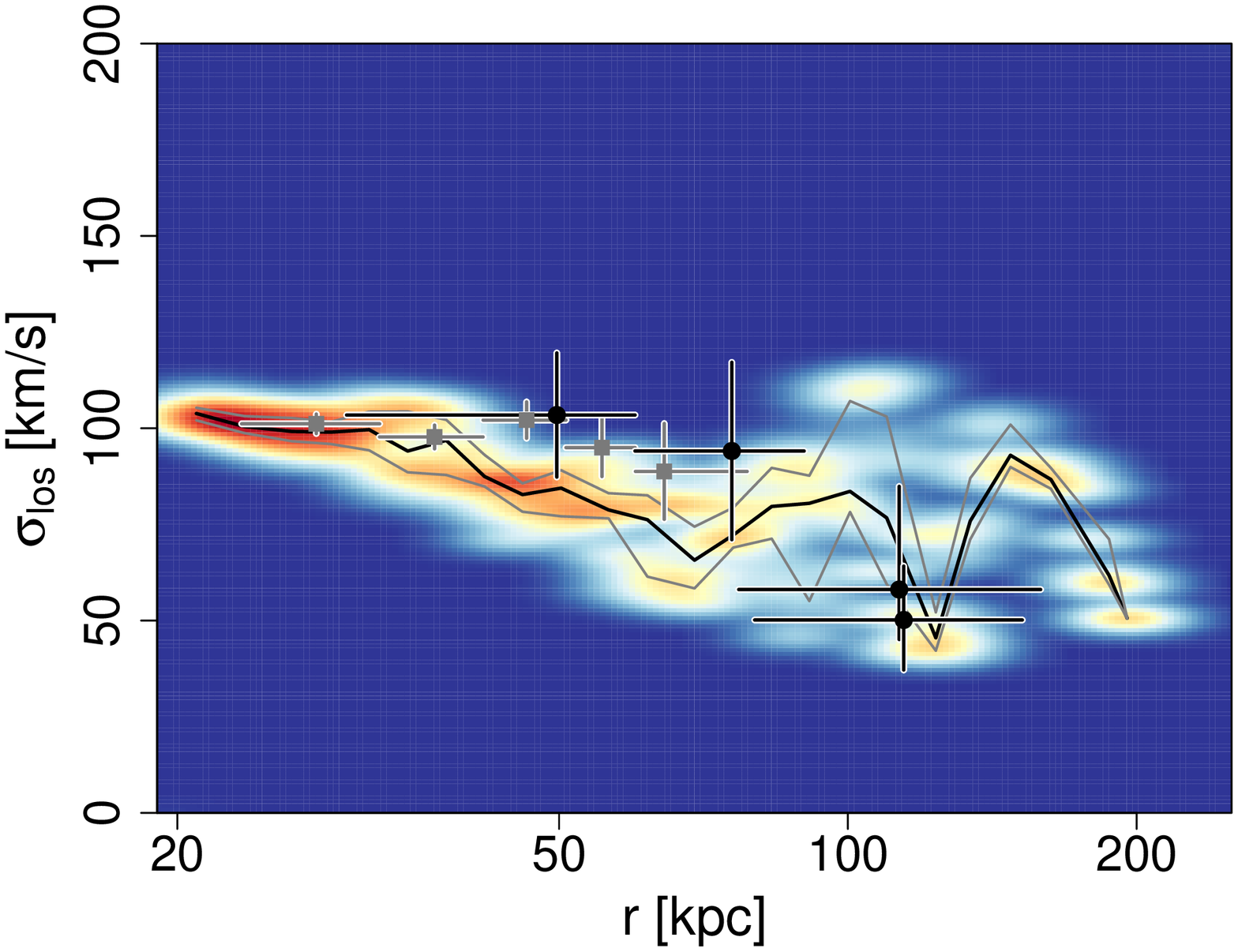}
\vspace{+0.cm}
\caption{Density profile and kinematic properties of Eris' stellar halo. Heatmaps are based on 500 realizations of a mock SDSS survey (see the text for details). 
We include only star particles with $r>20$ kpc to avoid contamination from the stellar disk and bulge. {Top panel}: stellar density profile as a function of galactocentric radius. 
{Middle panel}: velocity anisotropy parameter. {Bottom panel:} line-of-sight velocity dispersion. The SDSS data points from \citet{xue11} ({gray squares}) and the distant halo star sample of \citet{dea12b} ({black dots}) are plotted for comparison. 
In each panel the black curve marks the median value of all realizations, while the gray curves mark the 25th and 75th percentiles.
The numbers next to the color bar in the top panel represent the percentage probability that a random measurement will fall in a given point of the $xy$ plane.
}
\label{figEheat}
\vspace{+0.cm}
\end{figure}

For a meaningful comparison with the observations, we defined Eris' stellar disk as the ``Galactic Plane", removed bound satellites, and created mock SDSS samples of halo stars by placing 500 hypothetical observers in Eris' disk at 8 kpc from the center and different azimuthal angles. Simulated star particles were selected in regions corresponding to the SDSS footprint, i.e.  at high Galactic latitudes ($|b|>30$ degrees). In each of these ``observational data sets", we measured the line-of-sight velocity dispersion, stellar density, and anisotropy profiles as a function of Galactocentric radius. 
The resulting heatmaps are shown in Figure \ref{figEheat}.

Eris' median stellar density profile can be described by a power law, $\rho_*\propto r^{-\gamma}$, with a slope that steepens from $\gamma=3.4$ in the inner, $20<r<60$ kpc, regions 
to $\gamma=4.5$ beyond 60 kpc. In the galactocentric radial range 5 to 40 kpc, \citet{bel08} estimated a slope $\gamma$ between 2 and 4, broadly consistent with our results \citep[see also][]{dea11b, ses11}. The stellar velocity anisotropy profile, shown in the middle panel of Figure \ref{figEheat}, implies an increasingly radially biased velocity ellipsoid with distance, with stellar orbits becoming purely radial ($\beta\rightarrow 1$) beyond 100 kpc. Observational constraints from the solar neighborhood, where line-of-sight velocities and proper motion measurements are available, find radially biased orbits with $\beta \sim 0.7$ \citep[e.g.,][]{smi09,bon10}, in good agreement with the values predicted by our simulation. At larger distances in the halo ($10 \lesssim r/\mathrm{kpc} \lesssim 30$), constraints based solely on line-of-sight velocities are less clear-cut, with radial, isotropic and tangential orbits all quoted in the literature  \citep[e.g.,][]{sir04, kaf12, dea12a}. However, these measures are likely affected by modeling assumptions and/or the presence of substructure (see e.g., \citealt{dea13b}).

Eris' line-of-sight velocity dispersion is shown in the bottom panel. The profile falls off at large distances, from about $100\,\kms$ at 20 kpc to below $70\,\kms$ at $r>60$ kpc, with a large variance in the unrelaxed outskirts. The data points in the figure mark the measured line-of-sight velocity dispersion profile of SDSS halo stars, and show that the ``light'', concentrated dark matter halo of Eris provides a good fit to the observations. Most of the mock SDSS realizations show the same ``cold veil'' recently observed in the distant stellar halo of the MW by \citet{dea12b}, with tracers as cold as $\sigma_{\rm los}\approx 50\,\kms$ between 100 and 150 kpc. \citet{bat05} first noted this decline in the radial velocity dispersion using BHB and red giant stars as well as distant satellite galaxies and globular clusters. The presence in Eris of such a cold radial velocity dispersion is not caused by the dominance of tangential motions, as stellar orbits are nearly radial beyond 100 kpc. Rather, it corresponds to a sharp fall-off of the density profile of tracer halo stars. 

\begin{figure}
\centering
\includegraphics[width=0.47\textwidth]{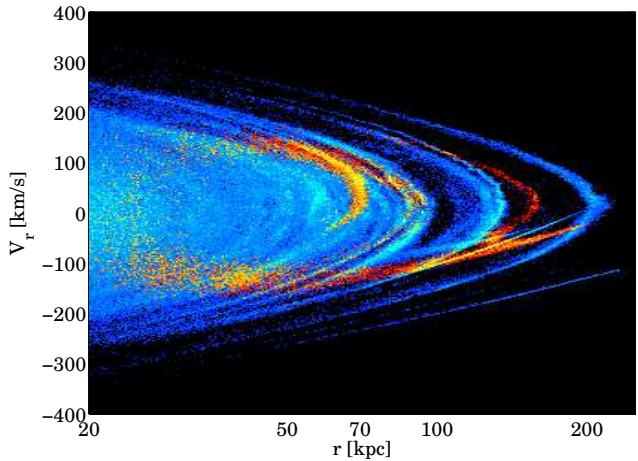}
\vspace{+0.cm}
\caption{Phase-space distribution of Eris' halo stars in $V_r$-$r$ space. The debris streams from a single accretion event -- an $M_*=1.7\times
10^8\,\msun$ satellite with an infall redshift of 1.2 -- are color coded (with the highest densities in yellow) for illustrative purposes.
}
\label{figphase}
\vspace{0.cm}
\end{figure}

The heatmaps in Figure \ref{figEheat} all reveal complex, structured behavior beyond 60 kpc, as expected from a hierarchically-formed stellar halo. While a 
detailed study of the evolution of substructure in phase-space is beyond the scope of this paper, here we show, in Figure \ref{figphase}, a two-dimensional 
radial velocity $V_r$ versus $r$ slice of all stellar tracers in Eris (excluding those still bound to satellites). Debris from the disruption of a satellite
phase mix over time along the progenitor's orbit while becoming locally colder in velocity space. The debris streams from a single accretion event -- an $M_*=1.7\times
10^8\,\msun$ satellite with an infall redshift of 1.2 -- are color coded for illustrative purposes. It is the overlapping of a few late events like this 
that causes the drop in the line-of-sight velocity dispersion around 70 kpc and the associated prominent ``dip'' in the $\beta$ parameter towards more isotropic stellar orbits. 
We note here that, over the last few years, several groups have found that the MW stellar halo has a ``broken'' density profile \citep[e.g.,][]{ses11,dea11b}, with a break radius of $\sim 30$ kpc and an inner slope of $\gamma=2.3-2.6$ steepening to $3.8-4.6$. \cite{dea13a} suggest that this break in the density profile of the MW stellar halo is due to a shell-type structure produced by an accretion event. Further support for this scenario comes from kinematic evidence of a ``cold'' feature in this radial regime (20-30 kpc) \citep[see][]{dea11a, kaf12, dea13b}. While the location of such features is clearly sensitive to the specific accretion history of the host galaxy, it is nevertheless noteworthy to identify similar components in Eris' halo.

\section{The Mass of the Milky Way's Dark Halo}
\label{mass_mass}

The results above appear to support a number of previous mass estimates based on SDSS halo stars, requiring a relatively light, centrally-concentrated Galactic dark matter halo. Our technique does not rely on the standard assumptions of spherical geometry, an NFW dark matter profile, a constant velocity anisotropy, and/or a scale-free tracer profile, and is based on a MW analog simulated in a fully cosmological context. Eris' total mass within 50 kpc is $M(<50)=3\times 10^{11}\,\msun$, entirely consistent with the value of $3.3\pm0.4\times 10^{11}\,\msun$ measured by \citet{dea12a} for a tracer density profile of slope 3.5. 

To better gauge the impact of a significantly more massive host on the predicted line-of-sight velocity dispersion of halo stars, we have performed a number of 
controlled experiments based on the integration of a simplified version of the Jeans equation. The assumption of a constant velocity anisotropy $\beta$ throughout the tracer sample 
results in the following solution for the radial velocity dispersion profile:
\begin{equation}
\sigma_r^2 = - {1\over r^{2\beta}\rho_*}\int_r^{\infty} dr' r'^{2\beta}\rho_* {GM(<r')\over r'^2}.
\end{equation}
\label{sigmar}
To proceed further, we have assumed a scale-free density profile for the stellar tracers, $\rho_*\propto r^{-\gamma}$, and numerically integrated the equation above by taking Eris' underlying 
gravity field and multiplying the mass inside some radius $r$ by a factor of 2.5. The last step is meant to mimic the potential of a $2\times 10^{12}\,\msun$ MW halo with the same ``scale
radius". 

\begin{figure}[th]
\centering
\includegraphics[width=0.47\textwidth]{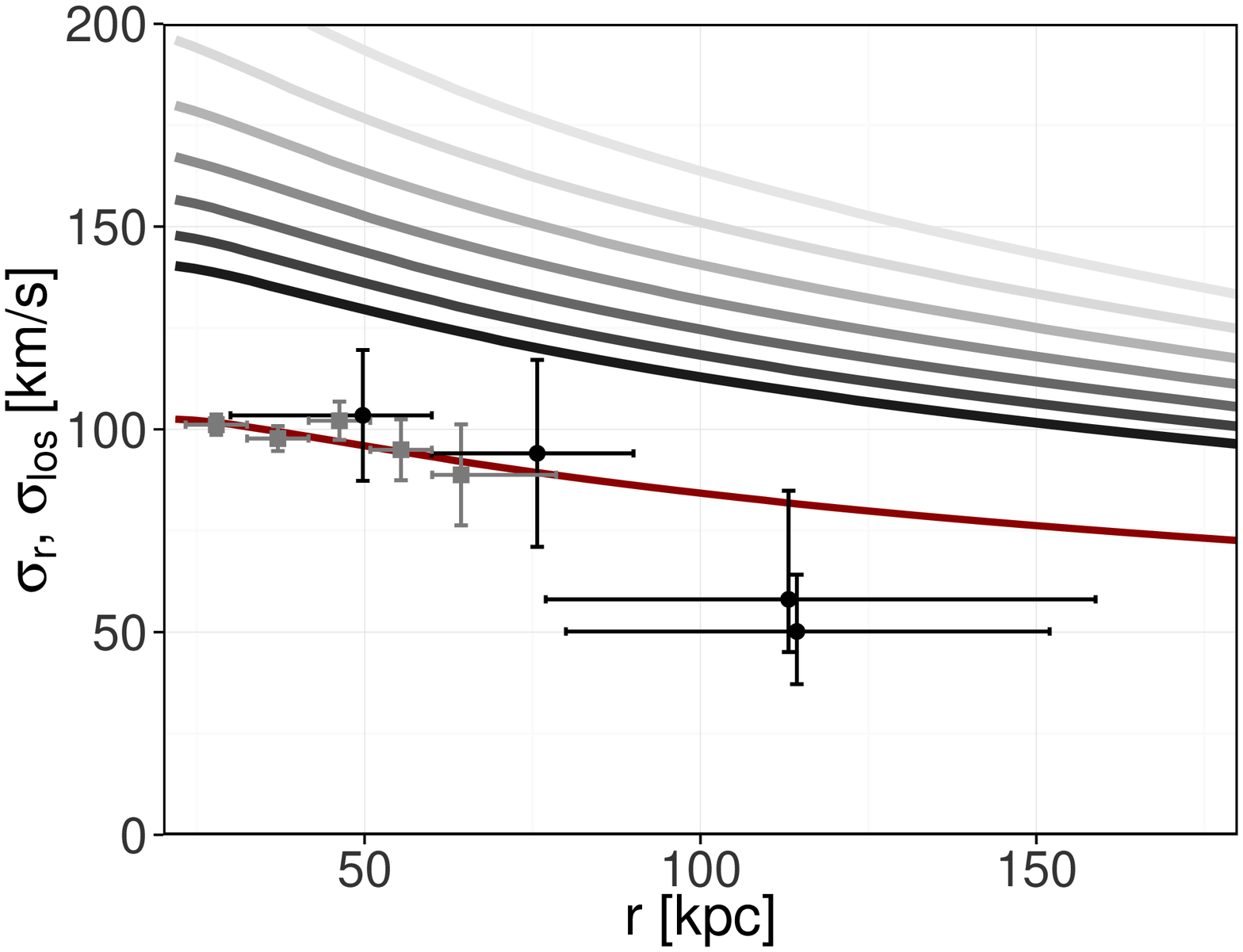}
\includegraphics[width=0.47\textwidth]{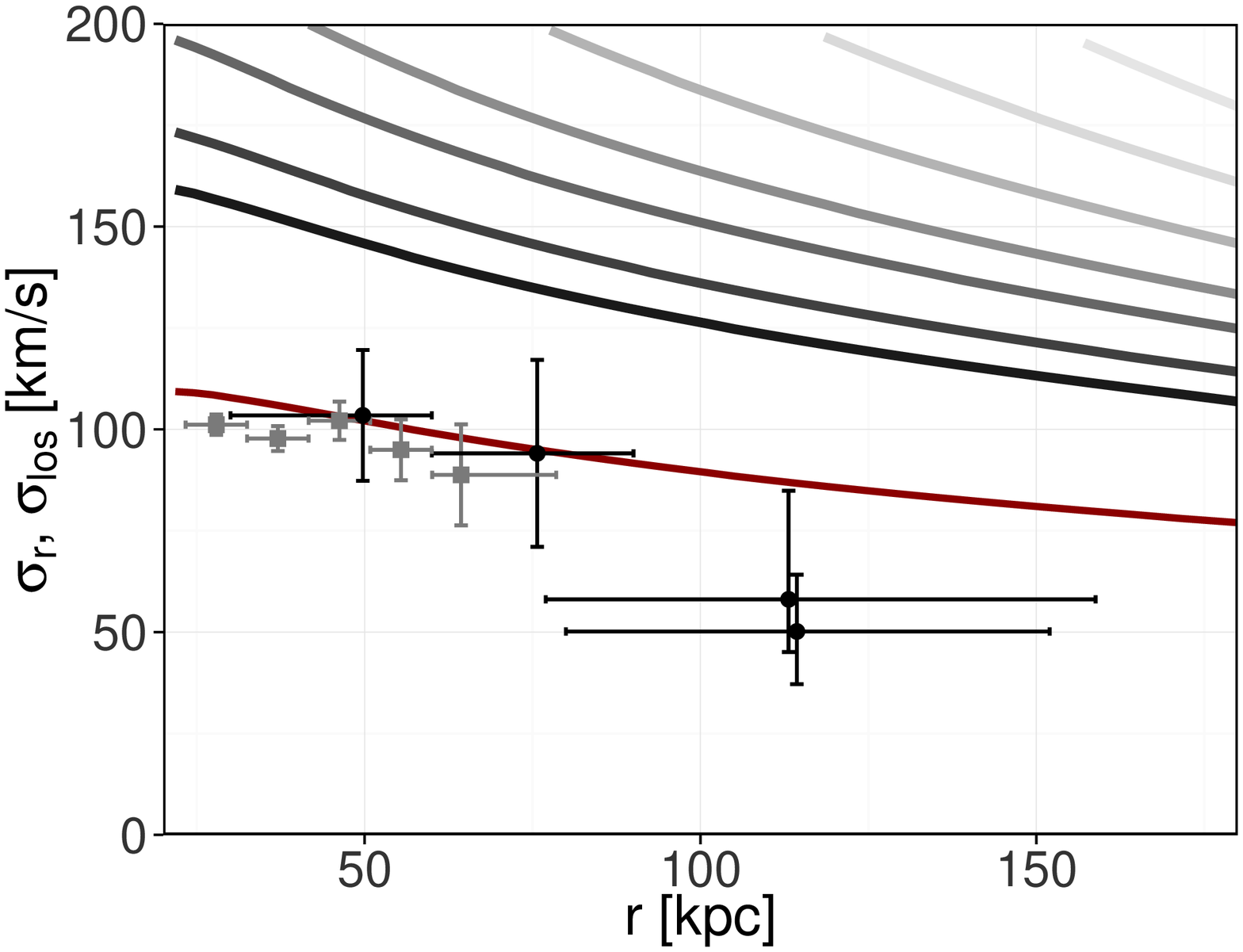}
\caption{\footnotesize The predicted radial velocity dispersion profile for a ``heavy''Eris halo (see the text for details).
The results of numerical integration of the Jeans equation, assuming a constant anisotropy $\beta$ and a scale-free stellar tracer profile $\rho_*\propto r^{-\gamma}$, are 
compared to the SDSS data. {Top panel:} $\gamma=3.4$ and $\beta=0.95, 0.75, 0.55, 0.35, 0.15, -0.05, -0.25$, and -2.0 (from top to bottom). 
{Bottom panel:} $\beta=0.75$ and $\gamma=1.5, 2.0, 2.5, 3.0, 3.5, 4.0, 4.5$, and 8 (from top to bottom). The SDSS data points are the same of Figure \ref{figEheat}.
}
\label{figjeans}
\vspace{+0.cm}
\end{figure}

Figure \ref{figjeans} show the results of such experiments, where the derived radial velocity dispersion is plotted against the line-of-sight velocity dispersion observed in SDSS halo stars. Note that, at large galactocentric distances, the line-of-sight velocity is almost identical to the radial velocity component, and $\sigma_r$ and $\sigma_{\rm los}$ can be directly compared without corrections. In the bottom panel, we have fixed the anisotropy $\beta$ to 0.75, and changed the slope of the tracer density profile from $\gamma=1.5$ to $\gamma=4.5$. The maroon line that provides a good fit to the data corresponds to a slope of $\gamma=8.0$! In the top panel, the slope of the tracer density profile is fixed to $\gamma=3.4$, and the anisotropy parameter ranges from $\beta=0.95$ to $\beta=-0.25$. 
Here, the maroon line that provides a good fit to the data corresponds to tangential orbits with $\beta=-2.0$. Clearly, only extreme values of $\beta$ and $\gamma$, values that are in contradiction with many observations, can accommodate a heavy Eris' halo. 

\begin{figure}
\centering
\vspace{+0.3cm}
\includegraphics[width=0.47\textwidth]{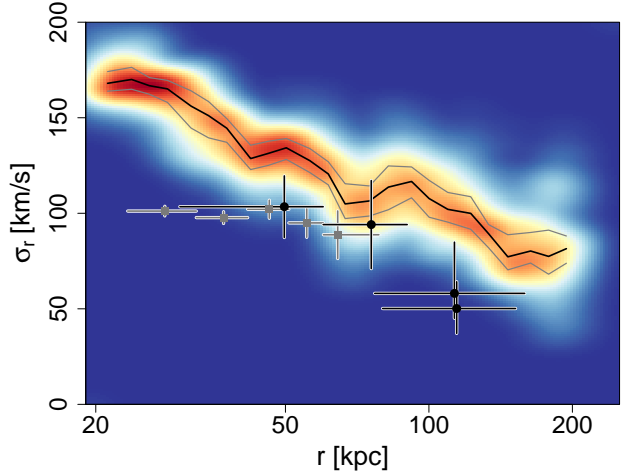}
\caption{Radial velocity dispersion profile for halo stars in the VLII simulation of a more massive host halo. Heatmaps are based on 500 realizations of a mock 
SDSS survey. The SDSS data points and the color coding are the same of Figure \ref{figEheat}. 
In each panel the black curve marks the median value of all realizations, while the gray curves mark the 25th and 75th percentiles.
}
\label{figVL2}
\vspace{+0.cm}
\end{figure}

Another example of the inability of a massive halo to reproduce the SDSS data is displayed in Figure \ref{figVL2}. Here, we have used the cosmological VLII simulation, one 
of the highest-resolution $N$-body calculations of the assembly of the Galactic halo to date \citep{die08}, and applied the particle tagging technique detailed in \citet{ras12} to 
dynamically populate it with halo stars. The method is calibrated using the observed luminosity function of MW satellites and the concentration of their stellar populations, and 
self-consistently follows the accretion and disruption of progenitor dwarfs and the buildup of the stellar halo in a cosmological ``live host''. VLII employs just over one billion 
$4,100\,\msun$ particles to model the formation of a ``heavy" $M_{\rm vir}=1.7\times \times 10^{12}\,\msun$ MW-sized halo and its satellites. As with Eris, we have selected 
500 mock SDSS surveys in VLII (where the lack of a stellar disk allows for complete freedom in choosing the orientation of the SDSS footprint), and produced heatmaps of the radial 
velocity dispersion profile. Compared to Eris, VLII is characterized by a steeper tracer density profile and a less radial anisotropy profile.
Figure \ref{figVL2} shows that the predicted radial velocity dispersion profile is a  
poor representation of the observations, and that VLII is simply too massive to support radial velocity dispersions of order $100\,\kms$ within the inner 100 kpc.

\section{Summary}
\label{mass_conclusion}

We have used a novel approach to ``weighing'' the mass of the MW halo, one that combines the latest samples of halo stars selected from the SDSS with
state-of-the-art numerical simulations of MW analogs. The fully cosmological, state-of-the-art runs employed in the present study include Eris, one of the 
most realistic hydrodynamical simulations of the formation of a $M_{\rm vir}=8\times 10^{11}\msun$ MW analog, and the dark-matter only $M_{\rm vir}=1.7\times 10^{12}\msun$ Via
Lactea II simulation. The kinematic properties of halo stars in the Eris simulation, supported by a relatively low-mass, highly concentrated 
dark matter halo, are in remarkably good agreement with the SDSS data. In particular, we find that the ``cold veil'' observed in the distant stellar halo of the MW by \citet{dea12b}, 
with tracers as cold as $\sigma_{\rm los}\approx 50\,\kms$ between 100-150 kpc, is also seen in most mock SDSS realizations in Eris. This feature is caused by the cold, 
unrelaxed regime of the stellar 
halo, where its radial extent approaches the virial radius of the galaxy. Using controlled experiments based on the integration of the spherical Jeans equation as well
as a particle tagging technique applied to VLII, we find that a ``heavy" MW host produces a poor fit to the kinematic 
SDSS data, unless the tracer density profile is very steep and/or the orbits of halo stars are tangentially biased. 

It is fair at this stage to caution the reader that, since in the outer halo regions of the simulation phase space is
not smoothly filled but structured according to a few random accretion events, the nice match 
between the observational data and the Eris simulation beyond 100 kpc may be the lucky outcome of a specific infall history. 
Still, our analysis shows that it would be much harder to match the observed radial velocity dispersion profile
in (say) the inner 70 kpc, where stars are more phase-mixed, with a more massive halo like VLII.
More Eris-like simulations, together with measurements of the tracer density profile and anisotropy over a 
large radial extent in the Galaxy using deeper photometric data (e.g., the Kilo Degree Survey) and proper motion 
surveys (e.g., from \textit{Gaia} and the {\it Hubble Space Telescope}), should all make a precise determination of the mass 
of the MW halo achievable in the near future.

\acknowledgments
P.M. acknowledges support by the NSF grant OIA-1124453 and the NASA grant NNX12AF87G.  
A.J.D. is currently supported by NASA through the Hubble Fellowship grant HST-HF-51302.01 awarded by the Space Telescope Science Institute, 
which is operated by the Association of Universities for Research in Astronomy, Inc., for NASA, under contract NAS5-26555. 
J.G. acknowledges support from the ETH Zurich Postdoctoral Fellowship and the Marie Curie Actions for People COFUND Program.

\end{document}